\newcommand{\QED}{\nopagebreak \hfill $\Box$}

\documentclass[12pt]{article}
\usepackage{amsmath, amssymb}

\usepackage{amsthm}

\usepackage{amsmath,amsthm}
\usepackage{multirow}
\usepackage{graphicx}
\usepackage{epstopdf}
\usepackage{ifthen}

\ifthenelse{\isundefined{\GeneratePdf}}
{
}
{
\usepackage[pdftex]{graphicx}
\usepackage[pdftex]{hyperref}
}

\usepackage{framed}
\ifthenelse{\isundefined{\NoGenerateLetterSize}}
{

\ifthenelse{\isundefined{\GeneratePdf}}
{}
{
\pdfpageheight\paperheight
\pdfpagewidth\paperwidth
}
}
{}

\long\def\LongVersion#1\LongVersionEnd{}
\long\def\ShortVersion#1\ShortVersionEnd{#1}

\newcommand{\Comment}[1]{}

\LongVersion %{
\LongVersionEnd %}

\ShortVersion %{
\ShortVersionEnd %}

\ShortVersion %{
\ShortVersionEnd %}

\newtheorem{theorem}{Theorem}[section]
\newtheorem{lemma}[theorem]{Lemma}
\newtheorem{observation}[theorem]{Observation}
\newtheorem{corollary}[theorem]{Corollary}
\newtheorem{proposition}[theorem]{Proposition}
\newtheorem{claim}[theorem]{Claim}

\theoremstyle{definition}
\newtheorem{property}[theorem]{Property}
\newtheorem{definition}[theorem]{Definition}
\theoremstyle{plain}
\newtheorem{fact}[theorem]{Fact}
\newtheorem{example}[theorem]{Example}

\theoremstyle{definition}

\theoremstyle{plain}
\Comment{
\newtheorem{theorem}{Theorem}[section]
\newtheorem{lemma}{Lemma}[section]
\newtheorem{corollary}{Corollary}[section]
\newtheorem{claim}{Claim}[section]

\newtheorem{definition}{Definition}[section]

\theoremstyle{definition}

\newtheorem{definition}[theorem]{Definition}
\theoremstyle{plain}
}

\newcommand{\Probability}[0]{\ensuremath{\hbox{\rm I\kern-2pt P}}}

\ShortVersion %{
\renewcommand{\paragraph}[1]{\par\noindent\textbf{#1}}
\ShortVersionEnd %}

\newcommand{\ignore}[1]{}
\usepackage{natbib}
\begin{document}
\title{Multi-Issue Social Learning}
\author{
Gal Bahar\thanks{
 Technion--Israel Institute of Technology, gal-bahar@campus.technion.ac.il}
 \and Itai Arieli\thanks{
 Technion--Israel Institute of Technology, iarieli@tx.technion.ac.il}
 \and Rann Smorodinsky\thanks{
 Technion--Israel Institute of Technology, rann@tx.technion.ac.il}
 \and Moshe Tennenholtz\thanks{
 Technion--Israel Institute of Technology, moshet@ie.technion.ac.il } }

\maketitle

\begin{abstract}
We consider social learning where agents can
only observe part of the population (modeled as neighbors on an undirected graph),
 face many decision problems, and arrival order of the agents is unknown.
The central question we pose is whether there is a natural observability graph
that prevents the information cascade phenomenon.  We introduce the `celebrities graph' and prove that indeed
it allows for proper information aggregation in large populations even when the order at which agents decide is random and even when different issues are decided in different orders.
\end{abstract}

\section{Introduction}

When making decisions on various topics, we often turn to see how our colleagues, friends and family members decide in similar circumstances. By doing so, we harness the collective knowledge with the hope of making better informed decisions. Such decisions pertain to  multiple issues, such as the choice of a restaurant, a mortgage plan, a service provider and so on. Anytime we face a new decision problem, we turn to look at those among our friends who have already faced a similar dilemma and have made a decision. As the order by which decisions are made differ from one issue to another, it may well be the case that for an individual agent, the relevant circle of influence changes from one issue to another.

Learning from what others do introduces an inherent trap, known as an `information cascade'. Under an information cascade, an initial set of agents is, due to sheer misfortune, ill-informed. As a result, these agents take an inferior action. Subsequent agents are then convinced that the aforementioned action is optimal (``how can so many agents be wrong?") and so dismiss their own private information and follow the herd by taking the inferior action as well.

The herding literature typically (and implicitly) assumes all agents are familiar with each other in the sense that an agent always observes the choice of action made by {\em all} its predecessors in this sequential decision-making process. In reality, however, this is hardly the case and we mostly view only a subset of predecessors. This turns out to be key in circumventing the information cascade problem.

Whereas partial observability seems to diminish the risk of an information cascade, it introduces a new difficulty to social learning. If people observe less peers, then there may be no hope for the collective knowledge to be aggregated and for learning to occur. Consider an extreme case, when agents never observe each other. Obviously no information cascades will occur yet no one can enjoy the collective knowledge. As a result, many agents are bound to take an inferior action. In this paper, we identify a natural observability graph that is sufficiently sparse to avoid information cascades but is also sufficiently dense to allow aggregation of information.

In fact, the observation graph we identify, allows for simultaneous learning over multiple issues. Each issue is associated with its optimal action and with a (random) order according to which people make their decision. The latter suggests some inherent difficulties: First, agents only observe part of their neighbors, those who preceded them in the realized order. They do not observe neighbors who decide later nor do they observe anything about the neighbors of their neighbors. For each of the neighbors they do observe, they only see her action and nothing else. Second, the same graph must be relevant to all issues and all realized orders. In particular, it cannot be designed in an online manner to fit the realized order at which agents choose (for more on this, see Section \ref{Sec:MultiVsSingleIssue}).%
\footnote{For a single issue and a deterministic order over the agents, it is quite trivial how to mitigate information cascades and guarantee learning with partial observability. We do so in the sequel. This design can be extended to a single issue with a random order if the graph can be constructed in an online manner.}

\subsection{Our contributions}

Our primary contribution is identifying the {\em celebrities graph} and showing its power in aggregating information for arbitrary sets of issues. The celebrities graph we introduce is a complete bi-partite graph with a small minority of the agents on one side (the `celebrities') and with the rest of the agents on the other side (the `commoners'). We use the term celebrities graph, as it is reminiscent of a social structure, where a small set of individuals, a priori similar to all other agents, emerge as opinion leaders (celebrities). These opinion leaders have a knack for distilling what the crowds will desire by monitoring many individuals and detecting early fads while the commoners primarily look up to these opinion leaders.

We provide three theoretical results regarding learning with the celebrities graph:
\begin{itemize}
\item
Our first result proves that learning is guaranteed with high probability when agents random order is drawn uniformly for each issue and independent across issues.
\item
Although a uniform distribution on the order of agents is quite natural (e.g., when each agent makes his move according to some Poisson clock), one can, of course, inquire as to whether a similar construct exists for arbitrary distributions. To this end, we prove that a version of the celebrities graph guarantees learning. The specific structure of the graph, in particular, which agents become celebrities and which commoners, depends on the details of those distributions.
\item
The previous observation builds on our third result. We show that a random version of the celebrities graph allows for learning even if the order at which agents make decisions is chosen in an adversarial manner. By this, we mean that an adversary can choose the decision-making order after learning the details of the observability graph.
\end{itemize}
The intuition behind the success of the celebrity graph to aggregate information is quite straightforward. With high probability, we expect a large number of commoners to arrive first. In such an event, the first celebrity to arrive observes a large independent sample and so is likely to take the optimal action. Any subsequent agent will be better off mimicking that celebrity.

We, furthermore, briefly discuss the difficulty in constructing an optimal observability graph (as opposed to $\epsilon$-efficient graphs for large societies). We show that this turns out to be a hard problem even when only a single issue is involved and the order by which decisions are made is known in advance!

Using simulations, we demonstrate the robustness of our theoretical results to `realistic' networks and small populations. By `realistic', we refer to preferential attachment networks. These are networks that form by attaching more vertices to a network where a new vertex's inclination to link with an existing one is a function of the popularity of the latter, as measured by the number of his neighbors. It has long been argued that the preferential attachment process forms a high fidelity model for various existing social networks, with the World Wide Web serving as a lead example \citep{Kleinbergbook}. Not only are preferential attachment graphs realistic, but they also induce graphs that are topologically close to our celebrities graphs with one major difference whereby celebrities are likely to connect with one another.%
\footnote{Adding links between celebrities in our celebrities graph will not affect the results reported here.}
Whereas our model assumes agents are Bayesian, we slightly depart from the Bayesian paradigm in our simulation (due to its computational complexity) and replace it with a simple majority rule (which is prevalent in the social learning literature, see
\citep{Laland2004}), whereby each agent follows the majority action while accounting both for the private information as well as neighbors' actions. This is a tractable approximation of the Bayesian optimal action each agent takes in our theoretical analysis.

Our simulations show that scale-free graphs do indeed exhibit efficient learning of optimal actions as compared with cliques; their efficiency levels converge to those obtained in the celebrities graph. In particular, this suggests that the asymptotic efficiency results we obtain are structurally robust.

We provide two possible ways to interpret the surprising benefits inherent in the celebrities graph---a positive interpretation and a constructive one.

\subsection*{A positive interpretation}

Is our celebrities graph a realistic depiction of actual social networks?
We  have already argued that the celebrities graph is an extreme variant of preferential attachment graphs which, in turn, are considered a good model for the graph structure in real scale-free networks. In this case, our results argue that for various networks, such as online social networks, one can expect social learning.

With yet some more imagination, one can possibly argue that the prevalence of celebrities---opinion leaders with no apparent quality to justify this position---is a social structure that has a very appealing property. Such a structure allows for efficient social learning.

\subsection*{A constructive interpretation}

Historically, online social networks have emerged spontaneously.  However, as more and more of the social interaction takes place online, these networks are now partly manipulated. For example, some of the algorithms of networks, such as Linkedin and Facebook, are responsible for those to whom we are introduced, while others prioritize the content we receive from those who are already friends.  More generally, in some cases, the set of agents observable to any particular agent is also manipulated by some central designer (or mediator). Thus, an alternative interpretation of our
results is in the framework of mechanism design, where a crucial part of the mechanism is the network topology.
Our results suggest how to structure social networks when social learning is a desired property and when the network structure cannot be dynamically adapted as issues arise.
%\footnote{Arguably, an alternative solution could have been to monitor and construct the observability edges in real-time, for each issue separately. By doing so one can ensure that the ad-hoc observability structure per issue has the guinea-pig structure and so entails learning.  This is hardly realistic as it requires real-time monitoring of issues and ad-hoc intervention of observability edges that may be computationally demanding and socially unacceptable. Recall that under a uniform distribution over agents' decision time the celebrities graph tolerates any set of issues that sporadically emerge.}

\subsection{Multi-issue and single-issue learning.}\label{Sec:MultiVsSingleIssue}

A key technical lemma in our analysis (Lemma \ref{lem:single issue}) shows that one can reduce any multi-issue problem to a single-issue problem (possibly with different parameters). This holds due to our underlying assumptions that the primitives of the different issues, such as the optimal alternative, the order over the agents and the signals, are independent across the issues. This begs the question of why we should study the multi-issue model to begin with and whether it has its own merit. Our reply to such a hypothetical question is twofold. First, it may not be obvious to the reader or to the practitioner that indeed the seemingly more complex setting reduces to the simpler one. Second, for a single-issue setting, one could argue that the network design need not take place ahead of time but rather in real time, as agents arrive. This leads to a very simple solution, which we mention in the paper, and dub the `guinea-pig' graph. However, for the multi-issue case, the transition from an offline network design to an online design does not seem so natural, as the learning of the various issues takes place in different orders and in different times and, thus, an equivalent online design is not guaranteed.

\subsection{Related literature}\label{sec:lit}

Our work is closely tied to an  exciting branch of the economics literature studying the celebrated
{\it herding} model~\citep{Banerjee, Bikhchandani}. This literature studies sequential information aggregation.
In this setting, agents arrive one after the other, possibly in some random order,  and each has to choose some action. The value of each action is the same for all
agents and depends on some underlying unknown state of nature. Each agent privately receives a stochastic signal (i.e., private information) that is correlated
with the state of nature. Thus, the choice of action for the agent arriving at stage $t$ is based on his signal as well as actions chosen (and not signals observed) by his predecessors. Using a similar framework to ours, the aforementioned papers and most of the follow-up literature study a setting where all agents can observe each other. The primary observation is that efficiency can only prevail, even when there is only a single issue, whenever signals are unbounded \citep{SmithSorensen}.
Signals are called bounded when the ex-post probability of the true state of nature conditional on the signal is bounded away from $1$, across all states and all signals. In contrast, we demonstrate an observability network that supports efficient learning of multiple issues even when signals are bounded.

Restricting agents' visibility in order to achieve efficiency was first proposed in \citep{Smithlphd}. Smith shows that if there is a single issue and a known order in which decisions are taken, then one can restrict the visibility of some agents (use them as guinea pigs) and enable efficient learning as the size of society grows. For Smiths' construction to work,  the number of guinea pigs grows to infinity with population size while its proportion shrinks to zero. This feature is employed in
\citep{sgroi2002} that calculates the optimal sample size of guinea pigs for any population size. The problem with such a guinea pig construction is that for it to work one must know the order at which agents make their decisions. Thus, when the structure is set in advance and the order is unknown, this construction will most likely fail.

\citep{Acemoglu} studies random evolution of observability graphs and provides sufficient conditions under which the number of guinea pigs in the realized graph grows to infinity while its proportion shrinks to zero (as in  \citep{Smithlphd}). \citep{Monzon2014} provides an example where limited observability and random arrival entails learning even without the existence of guinea pigs.
Note that \citep{Monzon2014}, in contrast with the previous papers, considers a model with unbounded signals.
\citep{Arieli2016} considers a herding model over the $m$-dimensional random lattice and show that as a function of the edge probability $p$ of the random lattice,
there exists a fixed proportion of agents $\rho^m_p$ who asymptotically observe an infinite number of isolated agents. \citep{Chierichetti_EtAl} study a design question where an observability graph is given but the designer has control over the order in which decisions are made. In their model, in addition to the informational externalities, there are also utilitarian externalities as the value of each action depends on the number of agents that adopt it.
Unfortunately, none of the aforementioned papers shed light on the existence of an efficient network with multiple issues.
A more general discussion of the herding topic, and many related pointers can be found in \citep{Kleinbergbook}.
%While, to the best of our knowledge, our work is the first to introduce the impact of the social visibility network in the context of herding,

The lion's share of use of a network structure in multi-agent AI papers with self-motivated agents is to model the effect of players' actions. In such models, an agent's utility is affected by both his own action as well as those taken by his neighbors (e.g. \citep{BrandtFHS09,AshlagiKT08,Alon2012}). In contrast, our work uses the network structure to model information externalities.

Our work complements works on sequential voting with complete information in the CS literature
\citep{XiaC10a,Elkind2010}.
Unlike the herding literature, in voting models, a multiplicity of issues was previously studied. \citep{XIACL11} study the case where multiple issues are presented to the voters sequentially and the voting procedure is done sequentially for each issue separately. It turns out that the order in which the issues are presented may affect the equilibrium outcome of the voting game.

An alternative model of social learning is when a population of agents act simultaneously and repeatedly, which is in contrast with our sequential setup. The centrality of a small subset of agents, those we refer to as `celebrities', plays a key role in those models. In fact, whereas centrality of a small set of agents drives our positive results, it appears as an inhibitor of learning in that setting. This is true when agents are rational and maximize expected payoff while resorting to the Bayes formula or when agents update an action in some ad-hoc non-Bayesian rule. In such models, the population may gravitate towards the action chosen by the celebrities and an initial bias of these celebrities then propagates to the whole population (e.g., \citep{Golubjackson,Mosselet}).

% Whereas in the voting literature agents decide make s social choice and so there is action externality among agents, in our model the only externality is with respect to the information. Whereas in the voting model each issue is voted on simultaneously, it is the sequential nature of decision making on each issue separately that drives our model.

\section{Model}

Consider a set of $N$ agents and $K$ decision problems, hereinafter referred to as {\it issues}. With each issue $k$ we associate a weight, $0< \lambda^k \leq 1, \sum_{k \in K} \lambda^k =1 $, which represents its social importance (we later on use this to formalize our objective function). For each issue, $k \in K$ an agent must choose one of two actions in the set $A^k=\{0,1\}$. The agents  share a common utility function $U=\sum_k U^k$, where $U^k:\Theta^k\times A^k \to {\mathbb{R}}$ with $\Theta^k=\{0,1\}$ being a binary set of states of nature. In particular, $U^k(\theta,a)=1$ whenever $\theta=a$, and $U^k(\theta,a)=0$ otherwise.%
\footnote{The results reported here are appropriate for agents using the more general weighted utility function, $U=\sum_k \lambda^k U^k$, with arbitrary weights, $\lambda^k>0$.}
In addition, each agent $i$ occupies a location (i.e., a node) in an undirected graph $G$ with a set of vertices $N$ and a set of edges $E$ which is called the \emph{social network.} Each such issue, $k\in K$, is associated with a random permutation, $\sigma^k:N \to N$, over the agents; this is the order at which agents take actions (decide) on that issue. Let $\sigma^k(i)$ denote the (random) stage in which agent $i$ is called to make a decision with respect to issue $k$. Let us denote by $Prob^k(\sigma)$ the probability of the permutation $\sigma \in \Sigma$, where $\Sigma$ denotes the set of all permutations.

The social learning game is played as follows:
For each issue $k$, the state of nature $\theta^k$ is drawn with equal probabilities and independently across issues. In addition, a permutation, $\sigma^k$ over the agents, is drawn according to some probability distribution. Time is discrete and at time $t$ agent $i$ must choose an action, $a^k \in \{0,1\}$, with respect to issue $k$ whenever $\sigma^k(i)=t$. Agents do not know the chosen states nor the permutations; however, each agent $i$ can observe a random signal $s^k_i \in S =\{0,1\}$ with $Prob(s^k_i=\theta^k|\theta^k)=0.5+\delta^k \ (0 < \delta^k < 0.5)$. Signals are drawn independently across agents and across issues conditional on the realized states $\theta^k$.
%\footnote{The assumption of independence across issues is not required for our results.}
In addition, before taking an action on issue $k$, every agent $i$ observes the actions taken by all agents $j$ that preceded him on that issue and are also observable to $i$
(formally, $\{j: \sigma^k(j)<\sigma^k(i) \wedge  (i,j)\in E \}$).
%\footnote{As the network structure and the permutation, $\sigma$, are common-knowledge, the assumption that agents know the predecessors' identity is %superfluous for our results. This, however, requires some modification of the celebrities graph to be described below (in particular we require %celebrities to observe each other) and some of the proofs.}

From here on, we will assume, for simplicity, that the signal structure is equal across issues. Formally, $\delta^k=\delta \ \ \forall k$.%
\footnote{Whenever the set of parameters, $\delta^k$, are uniformly bounded away from 0 and 0.5,  our results can be accomplished, with an adjustment of the size of population.}
%Thus, issues now differ only with respect to the order at which agents decide. We can now associate an issue with an order and consider the set of agent permutations as the set of potential issues.

%\begin{itemize} \item $\Theta$ is finite set of states of nature. \item $P \in \Delta(\Theta)$ is a prior distribution over the state space. We assume throughout that $P$ has full support. \item $N$ is a finite set of agents (at the risk of some confusion we often use $N$ to refer to the cardinality of this set as well). \item $\sigma$ is a random permutation of the set of agents. \item $A$ is a finite set of actions available to each agent. \item $S$ is a finite set of signals \item $G=(G_{\theta})_{\theta \in \Theta}$ where $G_{\theta} \in \Delta(S)$ is a probability distribution over the set of signals, $S$, corresponding to the state $\theta$. Conditional on the state $\theta$ all agents receive an independent noisy signal according to $G_{\theta}$. \item $U:\Theta\times A \to R$ is the common utility function for all agents \end{itemize}

%A state of nature in $\theta \in \Theta$ is chosen randomly from a known prior distribution with full support. In a society of $N$ agents, each receives some signal $s \in S$ drawn from the probability distribution $G_{\theta}$, independently from signals received by other agents. The agents deciding order is according to the random permutation $\sigma$ and must choose an action in $A$.

Note that a social network induces a game of incomplete information among the agents. In the {\emph Bayesian equilibrium} of the game, each agent maximizes the expected utility subject to her information. We assume for simplicity that if in equilibrium agent $i$ is indifferent between the two actions, then she takes the action that is identical to her private signal.

Our formal objective is to identify an observability graph $E$ that may depend on the population size and various other primitives of the model (e.g., the set of issues, the signal structure, the distributions over orders)  that is {\it $\epsilon$-efficient}:
\begin{definition}
The Observability graph $(N,E)$ is called  {\em $\epsilon$-efficient} if
$$
\forall \vec{\theta}=(\theta^1, \theta^2 \ldots \theta^K)\in \{0,1\}^k:
$$
$$
\sum_{k \in K} \lambda^k  \sum_{\sigma \in \Sigma}  Prob^k(\sigma) \frac{\sum_{n=1}^N  Prob(a_n^k=\theta^k|\sigma, \vec{\theta} )}{N} > 1-\epsilon.
$$
\end{definition}

In words, we consider the expected proportion of individuals to make a correct decision on issue $k$ and require that their weighted average be larger than $1-\epsilon$.

Is there a social network that could mitigate information cascades for a given issue? Is there one which could mitigate this for multiple issues? To demonstrate the subtlety of the question, let us consider two extreme network structures, where $E=N\times N$ or
$E=\emptyset$. We argue that both are not $\epsilon$-efficient for small enough $\epsilon$ and for any issue, no matter in what order the agents' decisions are made. The fact that  $E=N\times N$ is
not efficient is a direct consequence of informational cascades and potential herding when signals are bounded (which is the case whenever the number of signals is finite as in our model), whereas the case for $E=\emptyset$ is straightforward.

An intuitive approach that could result in positive results is a hybrid of the two aforementioned approaches---to allow some agents to choose actions
independently by restricting them to not observe anyone else (to make a decision in isolation). Once sufficient observations have been
made, the optimal action is identified with high probability. We refer to the set of agents that decide in isolation as `guinea pigs'. Assume that for a given issue the initial set of decision makers do not observe each other and so coincide with the set of guinea pigs, while all the followers can observe the guinea pigs. With high probability a majority of the guinea pigs will take the optimal action and all the followers can copy that action and so social learning prevails. However, if the set of guinea pigs does not coincide with the set of first movers (which is highly likely when the order is random) then such a network structure will fail on the corresponding issue. To overcome this difficulty with the guinea pig structure, we propose the following network structure. We will then argue that this structure induces a de-facto guinea pig structure for most arrival orders. This,  in turn, will yield the desired efficiency result.

\begin{definition}
A graph $G=(V,E)$ is called a {\it bi-clique}  if there exists a partition  $V_1,V_2 \subset V$%
\footnote{$V_1,V_2 \subset V$ is a partition if $V=V_1\cup V_2$, $V_1 \cap V_2 = \emptyset$.} 
and $E=\{(v,u) | v\in V_1, u \in V_2 \}$.%
\footnote{In other words, it is a bi-partite graph containing all edges between $V_1$ and $V_2$.}
\end{definition}

Whenever $|V_1| << |V_2|$ we shall informally refer to such bi-cliques as {\it celebrity graphs}, where  the vertices in $V_1$ are refered to as {\it celebrities} and those in $V_2$ as {\it commoners}.

%\end{definition}
%\begin{definition}
%The `celebrities social network' is a bi-clique graph $G=(V,E)$: $V=V_1\cup V_2$ where $V_1$ represents the "celebrities" $V_2$ represents the "commoners" and the following exists:
%\begin{itemize}
%\item $|V_1| = M << |V_2| = n-M$.
%\item $\forall (v_i\in V_1, v_j\in V_2): (v_i,v_j)\in E$
%\item $\forall (v_i\in V_1, v_j\in V_1): (v_i,v_j)\notin E$ \footnote{a similar analysis is all celebrities connect to one another, but commoners are not connected}
%\item $\forall (v_i\in V_2, v_j\in V_2): (v_i,v_j)\notin E$.
%\end{itemize}
We shall denote by ${\mathcal H}(N,M)$ the family of bi-cliques over the set of $N$ agents, where $|V_1|=M$ and $|V_2|=N-M$. Note that the identity of the celebrities changes from one graph in ${\mathcal H}(N,M)$ to another.

\section{Results}

Our main contribution is to show that celebrity networks admit an efficient outcome. We show this under three distinct arrival regimes. 

\begin{definition}
The {\it uniform arrival order} is the regime when all arrival orders are equally likely. Formally, when $prob^k(\sigma)=\frac{1}{N!}$ for all issues $k$ and all permutations $\sigma$.
\end{definition}

\begin{theorem} \label{THM101}
Let $\sigma$ be the uniform arrival order. For any $\epsilon > 0$, there exists $\ M=M(\epsilon)$ and $\hat{N}=\hat{N}(\epsilon)$ such that for any
$N>\hat N$ every celebrities graph in ${\mathcal H}(N,M)$ is $\epsilon$-efficient.
\end{theorem}

The second regime is an arbitrary random arrival order.

%\begin{definition}
%The {\it arbitrary arrival order} is the regime when arrival orders are arbitrary but the chance for each order is known. Formally, when $prob^k(\sigma)=p'^{K}(\sigma)$ for all issues $k$ and all permutations $\sigma$ and the probabilities $p'^{K}(\sigma)$ are all known.
%\end{definition}

\begin{theorem}\label{THM102}
Let $\sigma$ be an arbitrary arrival order. For any $\epsilon > 0$, there exists $\ M=M(\epsilon)$ and $\hat{N}=\hat{N}(\epsilon)$ such that for any
$N>\hat N$ there exists a celebrities graph ${\mathcal H}(N,M)$ which is $\epsilon$-efficient.
\end{theorem}

Note that the uniform distribution renders any celebrities graph efficient. This means that any subset of agents of size $M$ can be designated as the set of celebrities. In contrast, the second theorem argues that the choice of celebrities actually depends on the distribution over the arrival orders.

The last regime we study is an adversarial one. We construct network structure that is robust to the actual choice of the (possibly random) arrival order. This result hinges on a random network, for which the choice of the set of agents chosen as  celebrities is random:  

%\begin{definition}
%The {\it adversarial arrival order} is the regime when arrival orders are decided by an adversarial.
%\end{definition}
\begin{definition}
A {\it random celebrity graph} is a distribution over the set  ${\mathcal H}(N,M)$, where each subset of size $M$ is equally likely to be the set of celebrities.
\end{definition}
\begin{theorem}\label{THM103}
For any $\epsilon > 0$, there exists $\ M=M(\epsilon)$ and $\hat{N}=\hat{N}(\epsilon)$ such that for any
$N>\hat N$ the random celebrities graph ${\mathcal H}(N,M)$ is $\epsilon$-efficient for any arrival order, $\sigma$.
\end{theorem}

For all three regimes, the required population size and the required number of celebrities are given by  $\hat{N}(\epsilon) = \frac{128}{\epsilon}ln\frac{4}{\epsilon}max(\frac {8(1+2\epsilon\delta)}{\delta^{2}\epsilon},\frac{4(1+2\epsilon\delta)}{\epsilon\delta^{2}(0.5-\epsilon)},\frac{16}{(0.5-\delta)\frac{\epsilon}{2}})$.

Before we turn to the proof of these theorems, we provide the underlying intuition.

{\bf Proofs intuition:} An important preliminary observation is that the multi-issue scenario can be reduced to a singl issue scenario. In other words, if a given graph is efficient under some arrival order regime for a single issue, it will also be efficient for any set of $K$ issues (Lemma \ref{lem:single issue}).

Now, with only a single issue in mind, we recall that for any graph in ${\mathcal H}(N,M)$ the set of celebrities is a small minority ($M << (N-M)$). Thus, when the arrival order is uniform, then a large number of commoners is likely to precede the first celebrity to arrive (Lemmas \ref{LEM2} and \ref{THM3}). Since those commoners do not observe each other, their action fully reveals their signal and so they serve as ad-hoc guinea pigs. The majority signal is very likely to agree with the state of nature  (Lemma \ref{LEM1}) and so reveal the optimal action. For this reason, any celebrity who, upon arrival, observes a large number of commoners will mimic this majority (Lemmas \ref{OBS:follow_majority}, \ref{LEM11}, \ref{OBS:follow_majority_2} and Corollary \ref{CRL12} ) and, in turn, any commoner who observes celebrities that preceded will most likely view a consensus action which he will also mimic (Lemma \ref{OBS:follow_majority_2}). This wraps up the proof of Theorem \ref{THM101}.

%When the arrival order is distributed uniformly those results will hold for every group of $M$ celebrities which will prove the first result (Lemmas \ref{lem:single issue}, \ref{Lem:single_unfirom_issue}, Corollary \ref{CLR1} and Theorem \ref{THM101}).

The proof of Theorem \ref{THM102}, regarding an arbitrary arrival, follows as a corollary from Theorem \ref{THM103} which argues that the random celebrities graph is $\epsilon$-efficient for any arrival order, in particular for an arbitrary arrival order, $\sigma$, specified in Theorem \ref{THM102}. This implies that, on average, a celebrities graph is $\epsilon$-efficient for $\sigma$ which, in turn, implies that it must be $\epsilon$-efficient for at least one of these graphs and the result follows.

The proof of Theorem \ref{THM103} is based on the observation that no matter in which order agents arrive, the probability distribution over the population of the random celebrities graph (the order in which celebrities and commoners arrive) is equal to the uniform arrival order for a fixed celebrities graph (Lemmas \ref{LEM4}, \ref{LEM5}, Corollary \ref{CLR2} and Theorem \ref{THM5}). Consequently, Theorem \ref{THM103} follows from Theorem \ref{THM101}

\section{Detailed proofs}
%For the uniform distribution, we construct a simple graph, whereas for the arbitrary permutations we prove that such a graph exists. However, we do not discuss how to construct such a graph given the weights and permutation distributions. As for the adversarial case, we do provide an explicit construction, albeit of a random graph. This is then used to prove existence for the arbitrary permutations scenario.

%The basic network structure we propose will also do well when we extend the model beyond the binary and beyond the symmetric case, although the exact size of the required population, $N$, as a function of $\epsilon$, the required proximity to efficiency, may change.%
%\footnote{By symmetry we refer to the symmetry of the prior, and that of the distribution of signals.}
Before turning to discuss observability graphs for multi-issue learning, we require a few auxiliary results proven from Lemma \ref{OBS:follow_majority} to Lemma \ref{lem:single issue}. We will later use the auxiliary results to prove our main results.

\subsection{Auxiliary results}
An important building block for our analysis is to understand how to construct an asymptotically optimal network for a single issue with a deterministic order. The proposed graph for this case is based on the aforementioned guinea-pig approach.
Assume agents must decide on a single issue and that, without loss of generality, agent $i$ decides at time $i$ ($\sigma(i)=i$). Consider a graph structure such that agents $\{1,\ldots,N\}$ are guinea pigs (they do not observe each other) and assume agent $N+1$ observes all the guinea pigs. Let $a_n$ denote the action of agent $n$.
Then:

\begin{lemma}\label{OBS:follow_majority} % gal was observation
If all agents are playing their best-response, then \\
$\sum_{n=1}^N a_n\geq \frac{N}{2}+1 \implies \ \ a_{N+1}=1$, \\
$\sum_{n=1}^N a_n \leq \frac{N}{2}-1 \implies \ \ a_{N+1}=0$ and otherwise  $a_{N+1}=s_{N+1}$.
\end{lemma}

In other words, whenever there is a clear majority for one action among the $N$ guinea pigs, agent $N+1$ ignores his own signal
and follows the majority.

%The proof of this observation is quite straightforward. It is omitted due to lack in space and will appear in the full paper.
We sketch the proof of this lemma which is quite straightforward: Note that a guinea pig takes action $\theta$ whenever
his signal is equal to $\theta$. Therefore, the condition $\sum_{n=1}^N a_n \geq \frac{N}{2}+1$ is the same as $\sum_{n=1}^N
s_n \geq \frac{N}{2}+1$.
This implies that  $\sum_{n=1}^{N+1} s_n > \frac{N+1}{2}$, no matter what the value of $s_{N+1}$ is. This means that the
majority of signals among agents $1,\ldots,N+1$ is $1$. Knowing this, the optimal action for agent $N+1$ is to take
action $1$. The second case follows symmetric arguments. As for the third case, note that the condition implies that
$s_{N+1}$ equals the (weak) majority signal among agents $1,\ldots,N+1$ and the conclusion follows.

\begin{lemma}\label{LEM1}
Assume that $E=\emptyset$ (all agents are guinea pigs) and let $X$ denote the number of agents that take the correct action. Whenever $N \geq N_1 = \frac {1+2\epsilon\delta}{\delta^{2}\frac{\epsilon}{2}}$, $Prob(X \ge \frac{N+1}{2}) > 1-
\frac{\epsilon}{2}$.
\end{lemma}

\textbf{Proof:}
Note that as $E=\emptyset$, the only information that agents have is their signal. Maximizing expected utility implies that each agent
takes the action equal to the signal received. Let $X$ be the random variable that counts the number of agents whose signal
equals the state of nature and note that $X$ has a binomial distribution with parameters
$(N,0.5+\delta)$. Therefore $E(X) =  N(0.5+\delta)$ and $Var(X)= N(0.5+\delta)(0.5-\delta)$.

\begin{equation}\label{eq1}
%\begin{aligned}
Prob(X < \frac{N}{2}+1) = Prob(E(X)- X > E(X)-(\frac{N}{2}+1)) \le
$$
$$
\le Prob(E(X)- X > E(X)-(\frac{N}{2}+1)) +
$$
$$
Prob(X-E(X) > E(X)-(\frac{N}{2}+1)) =
$$
$$
= Prob(|E(X)- X| > E(X)-(\frac{N}{2}+1)).
%\end{aligned}
\end{equation}

Applying Chebyshev inequality:
\begin{equation}\label{eq2}
%\begin{aligned}
Prob(|E(X)- X| > E(X)-(\frac{N}{2}+1)) \le  \frac {Var(X)}{(E(X)-(\frac{N}{2}+1))^{2}} =
$$
$$
= \frac {N(0.5-\delta)(0.5+\delta)}{(N(0.5+\delta)-\frac{N}{2}-1)^{2}} =
 \frac {N(0.5-\delta)(0.5+\delta)}{(N\delta-1)^{2}}
 $$
 $$
  \frac {N(0.5-\delta)(0.5+\delta)}{N^2\delta^2-2N\delta +1} \leq \frac {N(0.5-\delta)(0.5+\delta)}{N^2\delta^2-2N\delta}
$$
$$
\frac {(0.5-\delta)(0.5+\delta)}{N\delta^2-2\delta} \leq \frac {1}{N\delta^2-2\delta}.
%\end{aligned}
\end{equation}

For $N \geq \frac {1+2\epsilon\delta}{\delta^{2}\frac{\epsilon}{2}}=N_1$ we get $\frac {1}{N\delta^2-2\delta} \le \frac{\epsilon}{2}$,
which, together with inequalities \ref{eq1} and \ref{eq2} prove the result. \QED
%Q.E.D
\\

The following corollary is immediate.

\begin{corollary}\label{CLR1}
For any $\epsilon >0$, there exists an integer $\hat N$ such that for any society with more than $\hat N$ agents and any
deterministic order, there exists an $\frac{\epsilon}{2}$-efficient guinea pig social network.
\end{corollary}

\textbf{Proof:}
Let $\hat N=\frac {1+2\epsilon\delta}{\frac{1}{16}\delta^{2}\epsilon^2}$ and let the first $K=\frac {1+2\epsilon\delta}{\delta^{2}\frac{\epsilon}{4}}$ agents that are making decisions
be the guinea pigs. Hence, from Lemma \ref{LEM1}, we get that with probability greater than $(1-\frac{\epsilon}{4})$ a clear majority of the guinea pigs take the optimal  action, $\theta$. By Lemma \ref{OBS:follow_majority}, we conclude that when such a clear majority emerges, all the remaining agents also take  the action $\theta$ regardless of their own signal. Denoting by $X$ the number of agents who take action $\theta,$ we get:
$$
\frac{E(X)}{N} > \frac{(N-K)(1-\frac{\epsilon}{4})}{N}  \ge  1-\frac{\epsilon}{2}
$$
where the last inequality follows from our choice of $\hat N$ and $K$. \QED

\begin{lemma}\label{LEM11}
Assume that $E=\emptyset$ (all agents are guinea pigs) and let $Y$ denote the number of agents that take the incorrect action. Whenever $N \geq N_2 = \frac {4}{(0.5-\delta)\frac{\epsilon}{2}}$, $Prob(Y \ge \frac{N(0.5-\delta)}{2}) > 1-
\frac{\epsilon}{2}$.
\end{lemma}

\textbf{Proof:}
Note that as $E=\emptyset$, the only information that agents have is their signal. Maximizing expected utility implies that each agent
takes the action equal to the signal received. Let $Y$ be the random variable that counts the number of agents whose signal
equals the wrong state of nature and note that $Y$ has a binomial distribution with parameters
$(N,0.5-\delta)$. Therefore $E(X) =  N(0.5-\delta)$ and $Var(X)= N(0.5+\delta)(0.5-\delta)$.

\begin{equation}\label{eq11}
%\begin{aligned}
Prob(Y < \frac{N(0.5-\delta)}{2}) = Prob(E(Y)- Y > E(Y)-(\frac{N(0.5-\delta)}{2})) \le
$$
$$
\le Prob(E(Y)- Y > E(Y)-(\frac{N(0.5-\delta)}{2})) +
$$
$$
Prob(Y-E(Y) > E(Y)-(\frac{N(0.5-\delta)}{2})) =
$$
$$
= Prob(|E(Y)- Y| > E(Y)-(\frac{N(0.5-\delta)}{2})).
%\end{aligned}
\end{equation}

Applying the Chebyshev inequality:
\begin{equation}\label{eq21}
%\begin{aligned}
Prob(|E(Y)- Y| > E(Y)-(\frac{N(0.5-\delta)}{2})) \le  \frac {Var(Y)}{(E(Y)-\frac{N(0.5-\delta)}{2}))^{2}} =
$$
$$
= \frac {N(0.5-\delta)(0.5+\delta)}{(\frac{N(0.5-\delta)}{2})^{2}} =
 \frac {(0.5-\delta)(0.5+\delta)}{\frac{N}{4}(0.5-\delta)^{2}}
 $$
$$
= \frac {(0.5+\delta)}{\frac{N}{4}(0.5-\delta)}
\leq \frac {1}{\frac{N}{4}(0.5-\delta)}.
%\end{aligned}
\end{equation}

For $N \geq \frac {4}{(0.5-\delta)\frac{\epsilon}{2}}$ we get $\frac {1}{\frac{N}{4}(0.5-\delta)} \le \frac{\epsilon}{2}$,
which, together with inequalities \ref{eq11} and \ref{eq21} prove the result. \QED
%Q.E.D
\\

The following corollary follows immediately from Lemma \ref{LEM11}
\begin{corollary}\label{CRL12}
Assume that $E=\emptyset$ (all agents are guinea pigs) and let $Y$ denote the number of agents that take the incorrect action. Whenever $N \geq N_3 = max(\frac {1+2\epsilon\delta}{\epsilon\delta^{2}(0.5-\delta)},\frac {4}{(0.5-\delta)\frac{\epsilon}{2}}) $, $Prob(Y \geq\frac{1}{2} \frac{1+2\epsilon\delta}{\epsilon\delta^{2}}) > 1-\frac{\epsilon}{2}$.
\end{corollary}

\textbf{Proof:}. Since $N \geq \frac {4}{(0.5-\delta)\frac{\epsilon}{2}}$ Lemma \ref{LEM11} implies that $Prob(Y \ge \frac{N(0.5-\delta)}{2}) > 1-
\frac{\epsilon}{2}$. However, since $N \geq \frac {1+2\epsilon\delta}{\epsilon\delta^{2}(0.5-\delta)}$ we get that
$Prob(Y \ge \frac{\frac {1+2\epsilon\delta}{\epsilon\delta^{2}(0.5-\delta)}(0.5-\delta)}{2}) > 1-
\frac{\epsilon}{2}$\\
 Hence $Prob(Y \geq\frac{1}{2} \frac{1+2\epsilon\delta}{\epsilon\delta^{2}}) > 1-\frac{\epsilon}{2}$
 \QED
%Q.E.D

We now turn to argue that our model with multiple issues can be reduced to a single-issue model with a random permutation on the order of agents. That is, given a setting with multiple issues, we construct an alternative setting with a single issue such that any $\epsilon$-efficient observability graph for the latter is necessarily  $\epsilon$-efficient for the former.

\begin{lemma}\label{lem:single issue}
Consider a set of $K$ issues with corresponding weights $\lambda^k$ and random orders with probabilities $Prob^k$. There exists a single-issue setting with a permutation over the agents such that if $E$ is $\epsilon$-efficient for the latter, then it is also $\epsilon$-efficient for the former. In particular, the  probability of the permutation $\sigma$ for the single issue is given by $q_\sigma = \sum_{k \in K} \lambda^k Prob^k(\sigma)$.
\end{lemma}

\textbf{Proof:}
Recall that the key term we compute to determine efficiency of $\vec{\theta}$ is:
$$
EFF(\vec{\theta}) = \sum_{k \in K} \lambda^k  \sum_{\sigma \in \Sigma}  Prob^k(\sigma) \frac{\sum_{n=1}^N  Prob(a_n^k=\theta^k|\sigma,\vec{\theta})}{N}.
$$

We can write this alternatively as:
$$
EFF(\vec{\theta})
=\sum_{k \in K} \lambda^k \sum_{\sigma \in \Sigma}  Prob^k(\sigma) \frac{\sum_{n=1}^N  Prob(a_{\sigma^{-1}(n)}^k=\theta^k|\sigma,\vec{\theta})}{N}.
$$

As all issues have the same parameters (prior and signal structure) then for any permutation $\sigma$, the set of probabilities
$\{Prob(a_{\sigma^{-1}(n)}^k=\theta^k|\sigma,\vec{\theta}): k \in K\}$ are all equal $Prob(a_{\sigma^{-1}(n)}=\theta|\sigma,\vec{\theta})$. Note that $Prob(a_{\sigma^{-1}(n)}=\theta|\sigma,\vec{\theta})$  is the probability of a successful decision of the agent deciding at stage $n$ in a model with a unique issue and unique permutation, $\sigma$.

Therefore
$$
EFF(\vec{\theta})
=\sum_{k \in K} \lambda^k \sum_{\sigma \in \Sigma}  Prob^k(\sigma) \frac{\sum_{n=1}^N  Prob(a_{\sigma^{-1}(n)}=\theta|\sigma,\vec{\theta)}}{N}=
$$
$$
=\sum_{\sigma \in \Sigma} \sum_{k \in K} [\lambda^k Prob^k(\sigma)] \frac{\sum_{n=1}^N  Prob(a_{\sigma^{-1}(n)}=\theta|\sigma,\vec{\theta})}{N}.
$$
For any $\sigma \in \Sigma$, let $q_\sigma = \sum_{k \in K} \lambda^k Prob^k(\sigma)$. Note that the vector $\{q_\sigma\}_{\sigma \in \Sigma}$ is a probability vector over the set of permutations. Thus,
$$
EFF(\vec{\theta})
=\sum_{\sigma \in \Sigma} q_\sigma \frac{\sum_{n=1}^N  Prob(a_{\sigma^{-1}(n)}=\theta|\sigma,\vec{\theta})}{N}.
$$

The latter term is the efficiency of the observability graph with a single issue and a random permutation distributed according to the probabilities $q_\sigma$. \QED

\subsection{Issues with a uniform distribution}\label{SE:Uniform}

In this subsection we focus on the case where for all issues the permutations have equal probabilities. Formally, $prob^k(\sigma)=\frac{1}{N!}$ for all issues $k$ and all permutations $\sigma$.

The idea we pursue is the following---designate agents $1,\ldots,M$ as `celebrities', where $M$ will be determined in the sequel (any subset containing $M$ agents would have worked as well). Let the agents in the complementary
set be known as `commoners'. The `celebrities graph' is a bi-clique with the $M$ celebrities on one side and the
$N-M$ commoners on the other side (a similar analysis is all celebrities connect to one another, but commoners are not connected).

For what follows, assume there is just one issue.

We now turn to show that for a random uniform order, all celebrities are likely to be absent from the set of agents that make decisions initially.
First we introduce the following notation: for any $M \subset N$ and a permutation $\sigma$, let $\sigma(M) = \{\sigma(m):m \in M \}$. Let $U$ be the uniform distribution over the set of permutations and denote by $\sigma_U$ the resulting random permutation.
\begin{lemma}\label{LEM2}
Fix two integers $M$ and $J$ and let $N$ be large enough to satisfy $N  \geq \frac{2JM}{\epsilon}+J$. %Let $\sigma_{U}\{1,\ldots,J\}$ denote for the agents arrived in places ($1, \ldots, J$),
Then any set of $M$ agents (also denoted $M$)
satisfies
$$
Prob(\{\sigma_{U}\{1,\ldots,J\}\cap M = \emptyset) \geq 1-\epsilon.
$$
%Prob(\{\sigma_{U}(1),\ldots,\sigma_{U}(J)\}\cap M = \emptyset) \geq 1-\epsilon.
%$$
\end{lemma}

\textbf{Proof:}
$Prob(\{\sigma_{U}\{1,\ldots,J\}\cap M = \emptyset) =$\\
$Prob(\sigma_{U}(1)\cap M = \emptyset)Prob(\sigma_{U}(2)\cap M = \emptyset|\sigma_{U}(1)\cap M =
\emptyset)$\\ $\ldots Prob(\sigma_{U}(J)\cap M = \emptyset|(\{\sigma_{U}\{1,\ldots,J-1\}\cap M = \emptyset)= $\\
%$ Prob(\sigma_{U}(1)\cap M = \emptyset)Prob(\sigma_{U}(2)\cap M = \emptyset|\sigma_{U}(1)\cap M =
%\emptyset)\ldots Prob(\sigma_{U}(J)\cap M = \emptyset|(\{\sigma_{U}(1),\ldots,\sigma_{U}(J-1)\}\cap M = \emptyset)=
$\frac {N-M}{N}  \frac
{N-M-1}{N-1} \ldots \frac {N-M-J+1}{N-J+1}
 > (\frac {N-M-J}{N-J})^{J}$.
Therefore, it is sufficient to show that $(\frac {N-M-J}{N-J})^{J} \geq 1-\epsilon$. However, this inequality is
equivalent to showing that $ N -J \geq \frac{M}{1- (1-\epsilon) ^{\frac{1}{J}}}$.
Recalling that $N-J  \geq \frac{2Jk}{\epsilon}$ it suffices to show that
$  \frac{2Jk}{\epsilon} \geq   \frac{M}{1- (1-\epsilon) ^{\frac{1}{J}}} $.

It is straightforward to verify that the following inequality holds: $e^{-2} \leq (1-\frac{1}{x})^{x} \leq e^{-1}$ for all $x \geq 2$.%
\footnote{The proof of this inequality follows from three simple observations: (1) $lim_{x\rightarrow \infty}(1-\frac{1}{x})^{x}=e^{-1}$, (2) $e^{-2} \leq (1-\frac{1}{2})^{2}$; and (3)  $(1-\frac{1}{x})^{x}$ is an increasing function for $x \geq 2$.}
By substituting  $\frac{2J}{\epsilon}$ for $x$ (note that indeed $ \frac{2J}{\epsilon} \geq 2$),
we have that $e^{-\epsilon} \leq (1-\frac{\epsilon}{2J})^{J}$.
As $1-\epsilon \leq e^{-\epsilon}$, we conclude that
$1-\epsilon \leq (1-\frac{\epsilon}{2J})^{J}$ or equivalently
$(1-\epsilon)^{\frac{1}{J}} \leq (1-\frac{\epsilon}{2J})$, which in turn implies that $\frac{M}{1 -
(1-\epsilon)^{\frac{1}{J}}} \leq \frac{M}{1-(1-\frac{\epsilon}{2J})}$. By simple manipulations, this implies that
$\frac{2JM}{\epsilon} \leq \frac{M}{1- (1-\epsilon) ^{\frac{1}{J}}}$ as required. \QED

Note that whenever $J$ is of the order of magnitude of $N\epsilon$ the conditions of Lemma \ref{LEM2} are violated. Indeed, it is quite likely that at least one celebrity will make a decision among the first $N\epsilon$ agents,
even when such celebrities are absent from the first $J$ agents:

\begin{lemma}\label{THM3}
For an arbitrary $\epsilon >0$, let $M,J$ and $N$ satisfy $M \geq \frac{2}{\epsilon}\ln(\frac{1}{\epsilon})$ and
$ N \geq \frac {2J}{\epsilon}$. % Let $\sigma_{U}\{1,\ldots,J\}$ denote for the agents arrived in places ($1, \ldots, J$) and $\sigma_{U}\{J+1,...,N\epsilon\}$ denote for the agents arrived in places ($J+1,...,N\epsilon$),
Then any set of $M$ agents (also denoted $M$)
satisfies  \\
$Prob(\sigma_{U}\{J+1,...,N\epsilon\}\cap M \not = \emptyset |
\sigma\{1,...,J\}\cap M = \emptyset) \geq 1-\epsilon.$
%$Prob(\{\sigma_{U}(J+1),\ldots,\sigma_{U}(N\epsilon)\}\cap M \not = \emptyset \ | \
%\{\sigma_{U}(1),\ldots,\sigma_{U}(J)\}\cap M = \emptyset) \geq 1-\epsilon.$
\end{lemma}

\textbf{Proof:}
$Prob(\sigma_{U}\{J+1,...,N\epsilon\}\cap M \not = \emptyset |
\sigma_{U}\{1,...,J\}\cap M = \emptyset) =$ \\
$1- Prob(\sigma_{U}\{J+1,..,N\epsilon\}\cap M = \emptyset |
\sigma_{U}\{1,..,J\}\cap M = \emptyset)$\\
%\begin{equation}\label{eq3}
%\begin{aligned}
% $Prob(\{\sigma_{U}(J+1),\ldots,\sigma_{U}(N\epsilon)\}\cap M \not = \emptyset \ | \
%\{\sigma_{U}(1),\ldots,\sigma_{U}(J)\}\cap M = \emptyset) =
%1- Prob(\{\sigma_{U}(J+1),\ldots,\sigma_{U}(N\epsilon)\}\cap M = \emptyset \ | \
%\{\sigma_{U}(1),\ldots,\sigma_{U}(J)\}\cap M = \emptyset) =
$ = 1 - \frac{N-J-M}{N-J} \frac{N-J-M-1}{N-J-1}... \frac{N-J-M-(N\epsilon - J)+1}{N-J-(N\epsilon - J)+1} \geq$\\
$ 1 - (\frac{N-J-M}{N-J})^{N\epsilon - J}.$
%\end{aligned}
%\end{equation}

Therefore, it is sufficient to show that $ (\frac{N-J-M}{N-J})^{N\epsilon - J} \leq \epsilon$.
As $ N \geq \frac {2J}{\epsilon}$, it is enough to show that
$(\frac{N-J-M}{N-J})^{\frac {N\epsilon}{2}} \leq \epsilon$.

Once again, we resort to the inequality  $ (1-\frac{1}{x})^{x} \leq e^{-1}, \ \forall x \geq 2$ and apply it to $x=\frac
{N-J}{M}$ (note that indeed $ \frac {N-J}{M} \geq 2$).
Thus $(\frac{N-J-M}{N-J})^{\frac {N\epsilon}{2}}\leq e^{-1\frac {N\epsilon}{2} \frac {M}{N-J}}\leq \epsilon$, where the
last inequality follows from our assumption that
$M \geq \frac{2}{\epsilon}\ln(\frac{1}{\epsilon})$. \QED

Given $\epsilon>0$ and a set of $N$ agents, let ${\mathcal H}$ be the bi-clique with
$M = \frac{8}{\epsilon}ln\frac{4}{\epsilon}$ celebrities and $N-M$ commoners (assume $N$ is large enough so that this is well defined).

\begin{lemma}\label{OBS:follow_majority_2}
For $J \geq J_1 = max(\frac {8(1+2\epsilon\delta)}{\delta^{2}\epsilon},\frac{4(1+2\epsilon\delta)}{\epsilon\delta^{2}(0.5-\epsilon)},\frac{16}{(0.5-\delta)\frac{\epsilon}{2}})$ and $\epsilon \leq (0.5-\delta)$. Assuming agents $1,\ldots, J$ are commoners, agent $J+1$ is a celebrity and agent $J+2$ is a commoner%
 \footnote{The same argument will hold when agent $J+2$ is also a celebrity}, assume all agents are playing their best reply. Then

\begin{enumerate}
\item $\sum_{h=1}^J a_n\geq \frac{J}{2}+1 \implies \ \   a_{J+2}=a_{J+1} \cap (w.p. > 1-\frac{\epsilon}{2}: \ a_{J+1}=1 ) \cap \forall J+3 <i <N: a_{i} = a_{J+1}$
\item $\sum_{h=1}^J a_n \leq \frac{J}{2}-1 \implies \ \ a_{J+2}=a_{J+1} \cap (w.p. > 1-\frac{\epsilon}{2}: \ a_{J+1}=0)\cap \forall J+3 <i <N: a_{i} = a_{J+1}$.
\end{enumerate}
\end{lemma}

\textbf{Proof:} First we will prove that agent $J+2$ necessarily mimics the action of agent $J+1$.  This is clear when $s_{J+2} = a_{J+1}$, namely the signal that agent $J+2$ receives is not in contradiction with the action taken by agent $J+1$. Thus, it remains to show that even when $s_{J+2} \not = a_{J+1}$ agent $J+2$ will choose the action $a_{J+1}$. Without loss of generality (WLOG) it is enough to prove this for the case $a_{J+1} = 1$ and $s_{J+2} =0$. Formally, we need to prove that   $Prob( \theta =1 | a_{J+1} = 1 \cap S_{J+2}= 0) >0.5$ whenever $J \geq J_1$.

By the Bayes rule  $Prob( \theta =1 | a_{J+1} = 1 \cap S_{J+2}= 0) = \\
\frac{Prob(\theta=1)Prob(a_{J+1}=1\cap S_{J+2} = 0|\theta=1)}{\sum_{j=0}^1 Prob(\theta=j)Prob(a_{J+1}=1\cap S_{J+2} = 0|\theta=j)}.$

Note that conditional on $\theta$ the random variables $a_{J+1}$ and $S_{J+2}$ are independent. In addition,
$Prob(S_{J+2} = 0|\theta=1)= 0.5-\delta$, $Prob(S_{J+2} = 0|\theta=0)= 0.5+\delta$ and from Lemmas \ref{OBS:follow_majority} and \ref{LEM1}  $Prob(a_{J+1}=1|\theta=1) \geq 1-\frac{\epsilon}{4} $ and $Prob(a_{J+1}=1|\theta=0) \leq \frac{\epsilon}{4} $ whenever $J \geq J_1$. Together with the trivial inequality $Prob(a_{J+1}=1|\theta=1) \leq 1$ and since $\epsilon \leq (0.5 - \delta)$ we can conclude that:

$$Prob( \theta =1 | a_{J+1} = 1 \cap S_{J+2}= 0) \ge $$
$$
\frac{0.5(1-\frac{\epsilon}{4})(0.5-\delta)}{0.5(1)(0.5-\delta) + 0.5\frac{\epsilon}{4}(0.5+\delta)} \geq $$
$$
\frac{(1-\frac{\epsilon}{4})(0.5-\delta)}{(0.5-\delta) + \frac{(0.5-\delta)}{4}(1)} =
\frac{(1-\frac{\epsilon}{4})}{\frac{5}{4}} = 0.8-\frac{\epsilon}{5} > 0.5.
$$

The proof that $a_{J+2}=a_{J+1}$ implies that once the first celebrity arrives, all the following commoners will mimic his action,  which implies that after the first celebrity arrives the minority of the actions will remain the same.  Corollary \ref{CRL12} implies that $w.p. > 1-\frac{\epsilon}{2}$ at least $\frac{1}{2} \frac{1+2\epsilon\delta}{\epsilon\delta^{2}}$ agents took the minority action. Hence, with probability greater than $1-\frac{\epsilon}{2}$, every celebrity can conclude that the first celebrity arrives after at least twice the number of commoners that took the minority action which is $\frac{1+2\epsilon\delta}{\epsilon\delta^{2}}$, and therefore, according to Lemmas \ref{OBS:follow_majority} and \ref{LEM1} together with the proof of the first part, the majority of commoners will point the correct action with probabiliy greater than $1-\frac{\epsilon}{2}$ and all the celebrities will follow it.  \QED
\begin{lemma}\label{Lem:single_unfirom_issue}
Let $\sigma$ be the uniform arrival order. For any $\epsilon > 0$, there exists $\ M=M(\epsilon)$ and $\hat{N}=\hat{N}(\epsilon)$ such that for any
$N>\hat N$ every celebrities graph in ${\mathcal H}(N,M)$ is $\epsilon$-efficient for a single issue.
\end{lemma}

\textbf{Proof:}
WLOG we can assume that $\epsilon \leq 0.5-\delta$. \\
Note that in the proof, we use Lemmas \ref{LEM2}, \ref{THM3}, \ref{OBS:follow_majority_2} and Lemma \ref{OBS:follow_majority} switching $\epsilon$ by $\frac{\epsilon}{4}$. \\
Let $J = max(\frac {8(1+2\epsilon\delta)}{\delta^{2}\epsilon},\frac{4(1+2\epsilon\delta)}{\epsilon\delta^{2}(0.5-\epsilon)},\frac{16}{(0.5-\delta)\frac{\epsilon}{2}})$, $\ M(\epsilon)=\frac{8}{\epsilon}ln(\frac{4}{\epsilon})$ and \\ $\hat{N}(\epsilon) = \frac{128}{\epsilon}ln\frac{4}{\epsilon}max(\frac {8(1+2\epsilon\delta)}{\delta^{2}\epsilon},\frac{4(1+2\epsilon\delta)}{\epsilon\delta^{2}(0.5-\epsilon)},\frac{16}{(0.5-\delta)\frac{\epsilon}{2}}) = 2\frac{8JM}{\epsilon} \geq \frac{8JM}{\epsilon} + J$.

Let A be the event that among the first $J$ agents who made a decision there is at least one agent from the set M. Let B be
the event that among the agents who made a decision in places $J+1$ to $N\frac{\epsilon}{4}$ there is no agent from the set M.

By Lemma \ref{LEM2}, $\ P(A)  \leq \frac{\epsilon}{4}$ which implies that $P(\bar{A}) \geq 1- \frac{\epsilon}{4}$. By Lemma
\ref{THM3}, $\ P(\bar{B}|\bar{A}) \geq (1-\frac{\epsilon}{4})$
and so $P(\bar{B}\cap\bar{A}) \geq (1- \frac{\epsilon}{4})^2$. \\
$ E(X) = P(A)(E(X)|A)+P(\bar{A})[P(B|\bar{A})(E(X)|B\cap\bar{A}) + P(\bar{B}|\bar{A})(E(X)|\bar{B}\cap\bar{A})]$ which
implies that
$ E(X)\geq P(\bar{B}\cap\bar{A})E(X|\bar{B}\cap\bar{A}) \ge (1- \frac{\epsilon}{4})^2 E(X|\bar{B}\cap\bar{A})$.

Assume we can show that  $E(X|\bar{B}\cap\bar{A}) \geq (1-\frac{\epsilon}{4})(N-N\frac{\epsilon}{4})$. Then we will be
able to conclude that $ E(X)\geq (1- \frac{\epsilon}{4})^4 N \ge (1- \epsilon) N$ as required.

To finish the proof, we now show that indeed
$E(X|\bar{B}\cap\bar{A}) \geq (1-\frac{\epsilon}{4})(N-N\frac{\epsilon}{4})$.  The event $\bar{A}$ implies that the initial set of $J= \frac {1}{\delta^{2}\epsilon}$ were all non-celebrities and hence
observe no predecessors. This means that they are all guinea pigs. Since $\bar{B}$ occurred, the first celebrity made a decision
prior to time $N\frac{\epsilon}{4}$ agents. This agent observed more than $J$ guinea pigs. By Lemma \ref{LEM1}, a strict
majority of these agents took the optimal action (their action equals their signal) with probability greater than
$1-\frac{\epsilon}{4}$. Therefore, by Lemma \ref{OBS:follow_majority}, the first celebrity followed this action. Any
agent joining after time $N\frac{\epsilon}{4}$ is either a non-celebrity, in which case he follows the action of the
first celebrity (follows from Lemma \ref{OBS:follow_majority_2}) or is himself a celebrity, in which case he sees
the same strict majority of initial $J$ commoners that he mimics. We conclude that conditional on the event
$\bar{B}\cap\bar{A}$, all agents making a decision after time $N\frac{\epsilon}{4}$ take an optimal action with
probability$1-\frac{\epsilon}{4}$ and our claim follows. \QED \\
Now, we are ready to prove Theorem \ref{THM101}.\\
\\
{\bf Theorem \ref{THM101}:} Let $\sigma$ be the uniform arrival order. For any $\epsilon > 0$, there exists $\ M=M(\epsilon)$ and $\hat{N}=\hat{N}(\epsilon)$ such that for any
$N>\hat N$ every celebrities graph in ${\mathcal H}(N,M)$ is $\epsilon$-efficient. \\
\\
\textbf{Proof:}
Recall that the term $q_\sigma = \sum_{k \in K} \lambda^k Prob^k(\sigma)$ introduced the reduction Lemma (Lemma  \ref{lem:single issue}). In the uniform setting, it is easy to see that $q_\sigma=\frac{1}{N!}$ which suggests that the reduction constructed in Lemma \ref{lem:single issue}  is for a single issue with a uniform order. The theorem now follows from Lemma \ref{Lem:single_unfirom_issue}.
\QED

\subsection{An adversarial order}

We now consider an adversarial setting regarding the weights assigned to the various issues and the corresponding order in which agents decide. As we show, we can resort to a random graph in order to obtain $\epsilon$-efficiency. The random graph we obtain will be central in proving the existence of a deterministic observability graph for any set of issues and random permutations. The random graph we pursue is, in fact, a random celebrities graph where the set of
celebrities, $M$, is not necessarily the set of agents $1,\ldots,M$ but is rather chosen randomly. The way we choose a
random celebrities graph is by randomly relabeling the agents and then applying the standard celebrities graph to the new
labeling---agents. Those labeled $1,\ldots,M$ are designated as celebrities and those labeled $M+1,\ldots,N$ are
designated as commoners.

More formally, let $\tau:N\to N$ be a random permutation of the names of the agents used by the designer (this is different from the order in which they made a decision). In particular let $\tau$ be similar to $\sigma_{U}$ in that it chooses each permutation with equal probability. Let ${\mathcal H}$ be the (random) celebrities graph where agent $k$ is a celebrity if and only if $\tau(k) \le M$.

Before considering the case of random relabeling, let us study the case where the relabeling is deterministic:

\begin{lemma}\label{LEM4}
Let $\tau$ and $\sigma$ be two deterministic permutations of the $N$ agents and consider the following two social
learning challenges which differ on the order by which decisions are made and corresponding social networks:
\begin{enumerate}
\item
Agents make decisions according to the order $\sigma$ with signals \\ %$\tau^{-1}(s_1),\tau^{-1}(s_2),\ldots,\tau^{-1}(s_N)$
$s_{\tau(1)},s_{\tau(2)},\ldots,s_{\tau(N)}$ and $\mathcal H$ is the bi-clique where agents
$\tau^{-1}(1),\ldots,\tau^{-1}(M)$ are designated as celebrities.
\item
Agents make decisions according to the permutation $\sigma(\tau^{-1})$ with signals $s_1,s_2,\ldots,s_N$
and ${\mathcal G}$ is the bi-clique with agents $1,\ldots,M$ designated as celebrities.
\end{enumerate}
For any signal profile $s_1,s_2,\ldots,s_N$ and for any stage $t$, the signal, the action and the location (celebrity or
not) of the agent making decisions at time $t$ in both challenges are the same.
\end{lemma}

\textbf{Proof:}
We prove this by induction. For $t=1$, the first agent to make a decision in $\mathcal H$ is $\sigma^{-1}(1)$ and he has the signal
$s_{\tau(\sigma^{-1}(1))}$. On the other hand, the first agent to make a decision in ${\mathcal G}$ is $\tau(\sigma^{-1}(1))$ which has
the same signal $s_{\tau(\sigma^{-1}(1))}$. Both have no additional information and so both also take the same action.
Also, in $\mathcal H$ agent $\sigma^{-1}(1)$ is a celebrity if and only if $\tau(\sigma^{-1}(1)) \le M$ which is exactly the
condition for agent $\tau(\sigma^{-1}(1))$, the one making a decision at time $t=1$ in $\mathcal G$, to be a celebrity.

Now assume the induction hypothesis holds for all agents making decisions at times $1,\ldots,t$.
Similar arguments as those made in $t=1$ show that the agents making decisions at time $t+1$ in both networks have the same signal
and are on the same side of the bipartite graph. By the induction hypothesis, the agents making decisions at time $t + 1$ also see the network spanned by the agents making decisions at times $1,\ldots,t$ (where the isomorphism function of agents is by the time they make a decision) and therefore they are isomorphic. In addition, the isomorphism saves the action profile. Hence, the agents making decisions at time $t$ have the same
additional information and so once again must take a similar action. \QED

Note that Lemma \ref{LEM4} is false if we were to assume that the signal vector in both networks is the same (e.g., had we replaced $s_{\tau(1)},s_{\tau(2)},\ldots,s_{\tau(N)}$ with  $s_1,s_2,\ldots,s_N$ in $\mathcal H$). However, as we now turn to show, whenever the signal vector is random, as in our model, the following holds.

\begin{lemma}\label{LEM5}
Let $\tau$ and $\sigma$ be two deterministic permutations of the $N$ agents and consider the following two social
learning challenges which differ on the order by which decisions are made and their corresponding social networks
\begin{enumerate}
\item
Agents make decisions according to the order $\sigma$, and $\mathcal H$ is the bi-clique where agents
$\tau^{-1}(1),\ldots,\tau^{-1}(M)$ are designated as celebrities.
\item
Agents make decisions according to the permutation $\sigma(\tau^{-1})$ and ${\mathcal G}$ is the bi-clique with agents $1,\ldots,M$ designated as celebrities.
\end{enumerate}
The expected number of agents taking the optimal action is the same, where the expectation is taken with respect to the random signal.
\end{lemma}

\textbf{Proof:}
Assume agents in $\mathcal H$ receive the signal vector $s$ while those in $\mathcal G$ receive the signal vector $\tau(s)$. From Lemma \ref{LEM4}, we conclude that the expected number of agents to take an optimal action in both cases is equal. In addition, note that any signal vector $s$ has the same probability as the signal vector $\tau(s)$ (this is the exchangeability property of conditionally independent signals). Therefore, as the function $s \to \tau(s)$ is one-to-one and onto the distribution over action profiles in $\mathcal G$ when agents receive the signal $s$ and when agents receive the signal $\tau(s)$. The claim now follows. \QED

As the above holds for any deterministic permutations, it must also hold for random permutations. Therefore we now have:

\begin{corollary}\label{CLR2}
Let $\tau$ and $\sigma$ be two random permutations of the $N$ agents and consider the following two social
learning challenges which differ on the order by which decisions are made and corresponding social networks
\begin{enumerate}
\item
Agents make decisions according to the order $\sigma$, and $\mathcal H$ is the bi-clique where agents
$\tau^{-1}(1),\ldots,\tau^{-1}(M)$ are designated as celebrities.
\item
Agents make decisions according to the permutation $\sigma(\tau^{-1})$ and ${\mathcal G}$ is the bi-clique with agents $1,\ldots,M$ designated as celebrities.
\end{enumerate}
The expected number of agents taking the optimal action is the same, where the expectation is taken with respect to the random signal and the two permutations, $\sigma$ and $\tau$.
\end{corollary}

Let us consider the special case where $\tau$ is chosen from the uniform distribution over the set of all permutations. In this case, we observe that the random permutation  $\sigma(\tau^{-1})$ is also uniform over all permutations. This implies that whenever agents make decisions according to the permutation $\sigma(\tau^{-1})$ and ${\mathcal G}$ is the bi-clique with agents $1,\ldots,M$ designated as celebrities, we are back to the case of Theorem \ref{THM101}.

We are now ready to prove the existence of a random celebrities graph that is $\epsilon$-efficient for a large enough society. Given $\epsilon>0$ and a set of $N$ agents, let ${\mathcal H}$ denote the random bi-clique where $\tau$ is chosen uniformly over all permutations and agent $n$ is a celebrity if and only if $\tau(n) \le M = \frac{8}{\epsilon}ln\frac{4}{\epsilon}$. Then:

\begin{theorem}\label{THM5}
For any $\epsilon > 0$, there exists $\hat N$, such that for any $N>\hat N$ the random network
${\mathcal H}$ is $\epsilon$-efficient for any single issue and any order $\tau$.
\end{theorem}
\textbf{Proof:}
WLOG assume $ \epsilon \leq (0.5-\delta)$. As the permutation $\sigma(\tau^{-1})$ chooses all permutations with equal probability, the claim follows from Theorem \ref{THM101} and Corollary \ref{CLR2}. \QED \\
We are now ready to prove Theorem \ref{THM103}\\
\\
{\bf Theorem \ref{THM103}: }
For any $\epsilon > 0$, there exists $\ M=M(\epsilon)$ and $\hat{N}=\hat{N}(\epsilon)$ such that for any
$N>\hat N$, the random celebrities graph ${\mathcal H}(N,M)$ is $\epsilon$-efficient for any arrival order, $\sigma$.
\\
\\
\textbf{Proof:}
This follows directly from Theorem \ref{THM5} and Lemma \ref{lem:single issue}. \QED

\subsection{Arbitrary permutations}

In the previous section, we applied the celebrities network in a random fashion to provide a social network that works
well for any (adversarial) order. However, let us consider, once again, the Bayesian setting where the order by which decisions are made is unknown but the distribution over orders is known. In contrast with Section \ref{SE:Uniform}, we will not assume uniform weights over all orders but an arbitrary vector of weights, $w_{\sigma}$.\\
We are now ready to prove Theorem \ref{THM102}\\
\\
{\bf Theorem \ref{THM102}: }
Let $\sigma$ be an arbitrary arrival order. For any $\epsilon > 0$ there exists $\ M=M(\epsilon)$ and $\hat{N}=\hat{N}(\epsilon)$ such that for any
$N>\hat N$, there exists a celebrities graph ${\mathcal H}(N,M)$ which is $\epsilon$-efficient.\\
\\
\textbf{Proof:}
By Theorem \ref{THM5}, we know that for $\epsilon >0$ there exists $\hat N$ such that for any $N>\hat N$ there exists a random graph that is $\epsilon$-efficient for a given issue and a random order of decision making. Recall the definition of an $\epsilon$-efficient network for a single issue requires that the expected proportion of agents taking an optimal action is larger than $1-\epsilon$. This can be computed as an expectation over the conditional expectation where the conditioning is over the realized graphs. This implies that at least one of these conditional proportions is larger or equal to $1-\epsilon$ and so the corresponding graph must be  $\epsilon$-efficient for the single issue. The theorem follows from applying the reduction lemma (Lemma \ref{lem:single issue}).%
\footnote{This result is in the spirit of the so-called probabilistic method \citep{ASE}.} \QED
%Q.E.D

\section{Preferential Attachment Networks}
The {\em celebrities graph} was shown to exhibit asymptotic efficiency. This settles the main foundational question we tackled in this paper. Given that, an interesting question is how do actual social networks, related to celebrities graph, perform. Naturally, such realistic networks exhibit a different structure and possibly different dynamics. In this section we report results from a simulation of social learning over {\em preferential attachment graphs} with a random order of agents that choose an action based on a naive {\em majority rule}. Given the reduction obtained in Lemma \ref{lem:single issue}, we suffice in demonstrating the performance in single-issue settings.

Preferential attachment graphs are arguably prevalent in reality (e.g., online social networks) \citep{Kleinbergbook}.
%GAL - WE NEED TO ADD SOME TEXT HERE ABOUT THE HISTORY OF THE MAJORITY RULE AND WHY IT APPROXIMATES THE BAYESIAN RULE.
A {\em Preferential attachment} graph is a graph that results from the following {\em preferential attachment} dynamics:
\begin{itemize}
\item The graph starts with $k$ nodes all connected to each other (the seed).
\item New nodes arrive sequentially and connect to some of the existing nodes.
\item Each new node is connected to $k$ nodes selected randomly, without returns, from the subsequent nodes, where the probability of node $i$, $p_i$,  is monotonic in the degree of $i$, denoted $d_i$.
\end{itemize}
We specifically study scale-free graphs for which the probability of connecting with each node is proportional to the power of the degree, $p_i = \frac{d_i^{\alpha}}{\Sigma_{j}d_j^{\alpha}}$. We denote such a graph by $PowPA(n,k,\alpha)$ or $PowPA(\alpha)$ when $n$ and $k$ are clear from the context. Note that as $\alpha$ grows, that graph approximates one where all agents connect only with all the seed nodes and so approximates a celebrities graph.  In addition, the difference between the described limiting graph and the celebrities graph is that in the former the celebrities are connected among themselves. The dynamics could diverge only when the first celebrity does not observe a clear majority and consequently follows his own signal. As this event occurs with low probability, the results obtained for the celebrities graph are also true for this limiting graph.

The case $PowPA(1)$ is known in the literature as the Barabasi-Albert preferential attachment graph
 \citep{Barabasi99} and was shown to approximate various real-life networks (e.g., the internet and citation networks) and in particular some social networks.
The dynamics we simulate are based on a majority rule and replace the Bayesian updating scheme underlying our theoretical results. This is done due to the complexity of implementation and running time.
  In majority dynamics, agents take the same action as the majority of their neighbors when also accounting for the optimal action implied from their private signal. We do so following a large body of work on emergent conventions with majority dynamics starting with \citep{STCONV}.
  Related dynamics have been discussed in \citep{Delgado}  for some preferential attachment graphs. The latter, however, refers to the stochastic pairing of agents in a game, rather than to social learning and efficient information aggregation. Notice that majority dynamics are close to the Bayesian dynamics we employed in our theory, and in graphs such as the celebrities graph.
\subsection{The mechanics of the simulation}
Given the parameters $n,k,\alpha$, we begin each Monte-Carlo simulation by generating a graph. We then randomly choose the state of nature with equal probabilities and finally proceed to simulate the agents, as follows:
\begin{itemize}
\item Each agent is associated with a node in the graph, and a random permutation over the agents is selected, determining their order.
\item For each agent we randomly choose a signal.
\item Each agent, in its turn, chooses an action according to a majority rule, where the majority is taken over the set of decisions made by the predecessors he observes (his neighbors on the graph) and his own signal, where the signal serves also as a tie-breaking rule.
\end{itemize}
For a fixed triplet of parameters  $(n,k,\alpha)$, we run many iterations of the simulation. The output we report---the success rate---is the average over the iterations of the proportion of agents whose action equals the state of nature.
\subsection{Simulation results}
Before we turn to report detailed results, let us note the two primary conclusions from the simulation. First, the efficiency results we get for Bayesian dynamics in the celebrities graph are robust in the sense that preferential attachment graphs with majority dynamics exhibit high efficiency. Second, the efficiency results are valid for surprisingly small populations  of agents -- as small as $n=1000$.

In the following three tables, we report the simulation results for  7 different graphs. Each table corresponds to a different  seed size and shows how the efficiency results vary with the signal parameter $p$. All the reported results are with respect to $10000$ iterations of the simulation.

%\begin{tabular}{ |p{4cm}||p{3cm}|p{3cm}|  }
% \hline
% \multicolumn{3}{|c|}{simulation results} \\
% \hline
% Graph name& number of simulations &Average Success Rate\\
% \hline
% Full Clique (1000)   & 100,000    &0.6915\\
% Celebrity(1000,30)&   100,000  & 0.906\\
% BA(1000,30)& 7,000  &   0.8427\\
% $PowPA(1000,30,\alpha=2)$    &10,000 & 0.8436\\
% $PowPA(1000,30,\alpha=3)$    &10,000 & 0.8864\\
% $PowPA(1000,30,\alpha=4)$    &10,000 & 0.8968\\
% $PowPA(1000,30,\alpha=5)$    &10,000 & 0.9049\\
 %$PowPA(1000,30,\alpha=6)$    &10,000 & 0.9008\\
 %$PowPA(1000,30,\alpha=8)$    &10,000 & 0.9022\\
 %$PowPA(1000,30,\alpha=10)$    &10,000 & 0.9088\\
 %$PowPA(1000,30,\alpha=12)$    &10,000 & 0.9016\\

% \hline
%\end{tabular} \\

\begin{center}
\begin{tabular}{ |p{1.2cm}|p{1.2cm}||p{1cm}|p{1cm}|p{1cm}|p{1cm}|  }
 \hline
% \rowcolor{green}
 \multicolumn{6}{|c|}{\bf {simulation results n=1000, k=20, 10,000 iterations}} \\
 \hline
 \multicolumn{2}{|c||}{\multirow{2}{*}{Graph}}& \multicolumn{4} {|c|} {Average Success Rate}\\
 \cline{3-6}
% \multicolumn{2}{|c||}{Graph}& \multicolumn{4} {|c|} {Average Success Rate}\\
 \multicolumn{2}{|c||}{}&p=0.6&p=0.7&p=0.8&p=0.9\\
 \hline \hline
 \multicolumn{2}{|c||}{Clique}   &0.6915  &0.8438 &0.9407 & 0.9876\\
 \hline
 \hline
 \multirow{5}{4em}{PowPA} & $\alpha=1$  &   0.8635 &0.9628& 0.9825 &0.9926\\
 &$\alpha=2$    & 0.8725 & 0.9623 & 0.9800 & 0.9913\\
 &$\alpha=3$    & 0.9047 & 0.9646 & 0.9809 &0.9906\\
 &$\alpha=4$     & 0.9108 & 0.9661 &0.9800 &  0.9904 \\
 &$\alpha=5$     & 0.9134& 0.9664 &0.9804&0.9906\\
 \hline \hline
 \multicolumn{2}{|c||}{Celebrity} & 0.9176 &0.9670 &0.9809 & 0.9910\\
 \hline
\end{tabular} \\
\end{center}

\begin{center}
\begin{tabular}{ |p{1.2cm}|p{1.2cm}||p{1cm}|p{1cm}|p{1cm}|p{1cm}|  }
 \hline
 %\rowcolor{green}
 \multicolumn{6}{|c|}{\bf {simulation results n=1000, k=30, 10,000 iterations}} \\
 \hline
 \multicolumn{2}{|c||}{\multirow{2}{*}{Graph}}& \multicolumn{4} {|c|} {Average Success Rate}\\
 \cline{3-6}
% \multicolumn{2}{|c||}{Graph}& \multicolumn{4} {|c|} {Average Success Rate}\\
 \multicolumn{2}{|c||}{}&p=0.6&p=0.7&p=0.8&p=0.9\\
 \hline \hline
 \multicolumn{2}{|c||}{Clique}   &0.6915  &0.8438 &0.9407 & 0.9876\\
 \hline
 \hline
 \multirow{5}{4em}{PowPA} & $\alpha=1$  &   0.8427 &0.9665& 0.9879 &0.9951\\
 &$\alpha=2$    & 0.8436 & 0.9639 & 0.9860 & 0.9942\\
 &$\alpha=3$    & 0.8864 & 0.9682 & 0.9855 &0.9938\\
 &$\alpha=4$     & 0.8968 & 0.9698 &0.9859 &  0.9935 \\
 &$\alpha=5$     & 0.9049& 0.9711 &0.9859&0.9934\\
 \hline \hline
 \multicolumn{2}{|c||}{Celebrity} & 0.906 &0.9716 &0.9862 & 0.994\\
 \hline
\end{tabular} \\
\end{center}

\begin{center}
\begin{tabular}{ |p{1.2cm}|p{1.2cm}||p{1cm}|p{1cm}|p{1cm}|p{1cm}|  }
 \hline
% \rowcolor{green}
 \multicolumn{6}{|c|}{\bf {simulation results n=1000, k=40, 10,000 iterations}} \\
 \hline
 \multicolumn{2}{|c||}{\multirow{2}{*}{Graph}}& \multicolumn{4} {|c|} {Average Success Rate}\\
 \cline{3-6}
% \multicolumn{2}{|c||}{Graph}& \multicolumn{4} {|c|} {Average Success Rate}\\
 \multicolumn{2}{|c||}{}&p=0.6&p=0.7&p=0.8&p=0.9\\
 \hline \hline
 \multicolumn{2}{|c||}{Clique}   &0.6915  &0.8438 &0.9407 & 0.9876\\
 \hline
 \hline
 \multirow{5}{4em}{PowPA} & $\alpha=1$  &   0.8295 &0.9634& 0.9901 &0.9963\\
 &$\alpha=2$    & 0.8300 & 0.9600 & 0.9879 & 0.9956\\
 &$\alpha=3$    & 0.8640 & 0.9651 & 0.9882 &0.9952\\
 &$\alpha=4$     & 0.8752 & 0.9676 &0.9874 &  0.9949 \\
 &$\alpha=5$     & 0.8799& 0.9720 &0.9880&0.9951\\
 \hline \hline
 \multicolumn{2}{|c||}{Celebrity} & 0.8903 &0.9718 &0.9887 & 0.9956\\
 \hline
\end{tabular} \\
\end{center}

Note that consistently the clique underperforms the preferential attachment graphs, which in turn, do not do as well as the celebrities graph. However, as $\alpha$ increases they do come very close. This observation is robust to the signal strength $p$ and the seed size $k$.

\section{Deterministic Arrival Order: Multi-layer Guinea Pigs}
Our work so far has focused on learning approximately optimal actions in large populations. This may not be of interest when considering small populations which begs the question of characterizing the optimal network, as opposed to showing an approximately optimal one. In this section, we demonstrate the underlying difficulty of this question by restricting the optimal design to a scenario where agents' order is well known.
In particular we examine the intuitive question of whether an optimal guinea-pig design is actually an optimal design and, therefore, the optimization problem boils down to identifying the optimal number of guinea pigs as a function of the population size. This last question has been studied in Sgroi  \citep{sgroi2002}.

Recall that Corollary \ref{CLR1} verifies the intuition that a guinea-pig design yields approximate efficiency for large enough populations. Unfortunately, as we now argue, despite this it is not the optimal design. In fact, simulation results show that for small populations ($n$=100) this is quite far from the first best design.

We now turn to introduce the ``multi-layer guinea-pigs" network which is instrumental for our argument:
\begin{definition}
Let $N=\{1,\ldots,n\}$ be the set of agents and assume that agents arrive according to the natural order. For any sequence of integers $0<n_1<n_2<\ldots<n_k \le n$ consider the graph  $G(n_1,n_2,\ldots,n_k)$ on $N$ with the following edges: $(v,w) \in E$ if and only if $\exists j$ such that $\min (v,w) \le \sum_{i\le j} n_i < \max (v,w)$.
\end{definition}
In other words, agents $1,\ldots, n_1$ are guinea pigs. Thereafter, each cohort of agents $n_i+1,\dots,n_{i+1}$ see the actions chosen by all previous cohorts but do not observe actions taken by their cohort peers. Finally, all agents from $n_k+1$ onwards observe all predecessors. In particular, the graph $G(n_1)$ is our standard guinea-pigs graph and so is a special case of a multi-layer guinea-pigs graph.
Fix a precision level $\delta>0$ for the signal distribution.
For every observation graph $G$ we let $X_G$ be the expected welfare, i.e., the expected number of agents that act correctly under the observation graph $G$.
For every number of agents $n$, let $s(n)$ be the optimal number of guinea-pigs
that should be sacrificed to maximize the welfare of the society in the guinea pigs graph. Let $X_n$ be the optimal welfare in the guinea0pigs graph as a function of $n$.
\begin{theorem}
For every large enough $n$ there exists a multi-layer graph $G=G(n_1,n_2,\ldots,n_k)$ such
that $X_n<X_G$.
\end{theorem}
\begin{proof}
Since, $\lim_{n\rightarrow\infty}\frac{X_n}{n}=1$, we clearly have that
$\lim_{n\rightarrow\infty}\frac{s(n)}{n}=0$. Therefore, we can choose a large enough $n_0$ such that $X_{n-s(n)}>(\frac{1}{2}+\delta)(n-s(n))$ for every $n>n_0$. Let $n_1=s(n),$ $n_2=s(n-s(n))$, and $G=G(n_1,n_1+n_2)$. We claim that $X_n<X_G$. To see this, let $p_1$ be the probability that after observing the first $n_1=s(n)$ agents, the remaining $n-s(n)$ assign the true state a probability that is strictly greater than $\frac{1}{2}+\delta$.
Similarly, let $p_2$ be the probability that after observing the first $n_1$ agents, the agents in $n-n_1$ assign the true state a probability smaller than $\frac{1}{2}-\delta$. Let $p_3=1-p_1-p_2>0$ be the remaining probability.
We can couple together the information structure in the guinea-pigs graph $G'=G(n_1)$ and in $G=G(n_1,n_1+n_2)$. In particular, we compare the  decision of the society in the two structures for identical realizations of signals. With probability $p_1$ in the two graphs, it holds that the $n-n_1$ remaining agents make the correct decision. This is clearly true for $G'$ and also for $G$ since if the agents in $n_2$ assign the state
$\theta$ a probability that is greater than $\frac{1}{2}+\delta$, then their optimal decision is to ignore their private information and follow the agents in $n_1$. Similarly, with probability  $p_2$ in the two graphs, the $n-n_1$ other agents make the incorrect decision.
The distinction between $G$ and $G'$ happens in the third case, where with probability $p_3$ the information from the first $n_1=s(n)$ make the other agents follow their private signal in $G'$. Since the other $n-n_1$ agents play according to their signal, their expected success probability is $(\frac{1}{2}+\delta)(n-n_1)$. By construction, the expected payoff of the other $n-n_1$ agents in $G$ is precisely $X_{n-n_1}>(\frac{1}{2}+\delta)(n-n_1).$ This shows that $X_n<X_G$. 
\end{proof}
Intuition suggests that the quantitative improvement of the multi-layer guinea-pig graphs over the single-layered one is insignificant for large populations. We now use simulations to demonstrate the improvement for intermediate populations sized ($n=30,50,70,100$). Each row compares the efficiency of two graphs: the optimal guinea-pig graph and the optimal multi-layered with layers of a constant size. The results reported all pertain to signals for which the precision is $p=0.6$. The average success rate is over 100000 iterations.
%As demonstrated in the proof of the theorem the multi-layer guinea pigs graph dominates the standard guinea pigs graph precisely when the group of guinea pigs is not decisive. While for a very large $n$ the welfare improvement is insignificant, for intermediate values of $n$, the welfare improvement may be more significant. To demonstrate this we use simulations for several population sizes $n$. For each $n$ we compare the optimal welfare in a guinea pig graph with the optimal welfare in a family of multi-layer graphs. We restrict attention to multi-layer graphs in which all layers are of the same size except for the last layer that may be of smaller size. In all cases we consider a signal precision of $p=0.6$. We evaluate the average success rate over 100000 iterations.\\
%
%Using naive simulation for multi-layer where all layers are the same size except for the last one which might be smaller \footnote{Note that this method is naive, and the optimal layer sizes shall be determined recursively. However this naive method already significantly improve the results comparing to one layer} we show the improvements in the following table: \\
\begin{center}
\begin{tabular}{ |p{2cm}||p{2cm}||p{2cm}|p{1cm}|  }
 \hline
 \multicolumn{4}{|c|}{simulation results, $p=0.6$} \\
 \hline
Population size& Graph type & Success Rate & Layer size\\
 \hline
30 & guinea pigs      &0.7028 & 10 \\
30 & multi-layer  & 0.7237 &  7\\
\hline
50 & guinea pigs       &0.7369 & 16\\
50 & multi-layer  & 0.7590 &   11\\
\hline
70 & guinea pigs      &0.7637 & 22 \\
70 & multi-layer  & 0.7845 &   15\\
\hline
100 & guinea pigs       &0.7902 & 30\\
100 & multi-layer  & 0.8107 &   21\\
\hline

\end{tabular}\\
\end{center}
~\\

Note that improvement rates are consistently around $3\% $. These would even increase if we move to multi-layer graphs with varying layer size.

\section*{Acknowledgments}\label{sec:Acknowledgments}
Itai Arieli gratefully acknowledges the support of the Ministry of Science and Technology grant 2025117.
Moshe Tennenholtz and Gal Bahar are funded by the European Research Council (ERC) under the European Union's Horizon 2020 research and innovation programme (grant agreement no.  740435). Rann Smorodinsky is funded by the joint United States-Israel Binational Science Foundation and National Science Foundation grant 2016734, German-Israel Foundation grant I-1419-118.4/2017,  Ministry of Science and Technology grant 19400214, Technion VPR grants, and the Bernard M. Gordon Center for Systems Engineering at the Technion.

\bibliographystyle{plainnat}
\bibliography{bibnew}

\begin{thebibliography}{23}
\providecommand{\natexlab}[1]{#1}
\providecommand{\url}[1]{\texttt{#1}}
\expandafter\ifx\csname urlstyle\endcsname\relax
  \providecommand{\doi}[1]{doi: #1}\else
  \providecommand{\doi}{doi: \begingroup \urlstyle{rm}\Url}\fi

\bibitem[Acemoglu et~al.(2010)Acemoglu, Dahleh, Lobel, and Ozdaglar]{Acemoglu}
Daron Acemoglu, Munther~A. Dahleh, Ilan Lobel, and Asuman Ozdaglar.
\newblock Bayesian learning in social networks.
\newblock \emph{The Review of Economic Studies}, 78:\penalty0 1--34, 2010.

\bibitem[Alon et~al.(1992)Alon, Spencer, and Erdos]{ASE}
N.~Alon, J.H. Spencer, and P.~Erdos.
\newblock \emph{The Probabilistic Method}.
\newblock John Wiley \& Sons, 1992.

\bibitem[Alon et~al.(2012)Alon, Babaioff, Karidi, Lavi, and
  Tennenholtz]{Alon2012}
Noga Alon, Moshe Babaioff, Ron Karidi, Ron Lavi, and Moshe Tennenholtz.
\newblock Sequential voting with externalities: Herding in social networks.
\newblock In \emph{ACM Conf. on Economics and Computation (EC)}, 2012.

\bibitem[Arieli and Mueller-Frank(2018)]{Arieli2016}
Itai Arieli and Manuel Mueller-Frank.
\newblock Multidimensional social learning.
\newblock \emph{The Review of Economic Studies}, 0:\penalty0 1--28, 2018.

\bibitem[Ashlagi et~al.(2008)Ashlagi, Krysta, and Tennenholtz]{AshlagiKT08}
Itai Ashlagi, Piotr Krysta, and Moshe Tennenholtz.
\newblock Social context games.
\newblock In \emph{WINE}, pages 675--683, 2008.

\bibitem[Banerjee(1992)]{Banerjee}
A.V. Banerjee.
\newblock A simple model of herd behavior.
\newblock \emph{The Quarterly Journal of Economics}, 107:\penalty0 797--817,
  1992.

\bibitem[Barabasi and Albert(1999)]{Barabasi99}
A.~Barabasi and R.~Albert.
\newblock Emergence of scaling in random networks.
\newblock \emph{Science}, 286:\penalty0 509�--512, 1999.

\bibitem[Bikhchandani et~al.(1992)Bikhchandani, Hirshleifer, and
  Welch]{Bikhchandani}
S.~Bikhchandani, D.~Hirshleifer, and I.~Welch.
\newblock A theory of fads, fashion, custom and cultural change as information
  cascade.
\newblock \emph{The Journal of Political Economy}, 100:\penalty0 992--1026,
  1992.

\bibitem[Brandt et~al.(2009)Brandt, Fischer, Harrenstein, and
  Shoham]{BrandtFHS09}
Felix Brandt, Felix~A. Fischer, Paul Harrenstein, and Yoav Shoham.
\newblock Ranking games.
\newblock \emph{Artif. Intell.}, 173\penalty0 (2):\penalty0 221--239, 2009.

\bibitem[Chierichetti et~al.(2012)Chierichetti, Kleinberg, and
  Panconesi]{Chierichetti_EtAl}
F.~Chierichetti, J.~Kleinberg, and A.~Panconesi.
\newblock {How to schedule a cascade in an arbitrary graph}.
\newblock In \emph{Proceedings of the 13th ACM Conference on Electronic
  Commerce}, pages 355--368, 2012.

\bibitem[Delgado(2002)]{Delgado}
Jordi Delgado.
\newblock Emergence of social conventions in complex networks.
\newblock \emph{Artif. Intell.}, 141:\penalty0 171�--185, 2002.

\bibitem[Desmedt and Elkind(2010)]{Elkind2010}
Yvo Desmedt and Edith Elkind.
\newblock Equilibria of plurality voting with abstentions.
\newblock In \emph{Proceedings of the 11th {ACM} Conference on Electronic
  Commerce (EC-2010), Cambridge, Massachusetts, USA, June 7--11, 2010}, pages
  347--356.

\bibitem[Easley and Kleinberg(2010)]{Kleinbergbook}
David~A. Easley and Jon~M. Kleinberg.
\newblock \emph{Networks, Crowds, and Markets - Reasoning About a Highly
  Connected World}.
\newblock Cambridge University Press, 2010.

\bibitem[Golub and Jackson.(2010)]{Golubjackson}
B.~Golub and M.~Jackson.
\newblock Naive learning in social networks and the wisdom of crowds.
\newblock \emph{American Economic Journal: Microeconomics}, 2:\penalty0
  112�--149, 2010.

\bibitem[Laland(2004)]{Laland2004}
Kevin~N. Laland.
\newblock Social learning strategies.
\newblock \emph{Learning and Behavior}, 32, 1:\penalty0 4�--14, 2004.

\bibitem[Monzo³n and Rapp(2014)]{Monzon2014}
Ignacio Monzo³n and Michael Rapp.
\newblock Observational learning with position uncertainty.
\newblock \emph{Journal of Economic Theory}, 154:\penalty0 375--402, 2014.

\bibitem[Mossel et~al.(2015)Mossel, Sly, and Tamuz]{Mosselet}
E.~Mossel, A.~Sly, and O.~Tamuz.
\newblock Strategic learning and the topology of social networks.
\newblock \emph{Econometrica}, 83:\penalty0 1755�--1794, 2015.

\bibitem[Sgroi(2002)]{sgroi2002}
Daniel Sgroi.
\newblock Optimizing information in the herd : guinea pigs, profits, and
  welfare.
\newblock \emph{Games and Economic Behavior}, 39,1:\penalty0 137�--166, 2002.

\bibitem[Shoham and Tennenholtz(1992)]{STCONV}
Y.~Shoham and M.~Tennenholtz.
\newblock {Emergent conventions in multi-agent systems: initial experimental
  results and observations}.
\newblock In \emph{Proc. of the 3rd International Conference on Principles of
  Knowledge Representation and Reasoning}, pages 225--231, 1992.

\bibitem[Smith and Sorensen(2000)]{SmithSorensen}
L.~Smith and P.~Sorensen.
\newblock Pathological outcomes of observational learning.
\newblock \emph{Econometrica}, 68:\penalty0 371--398, 2000.

\bibitem[Smith(1991)]{Smithlphd}
Lones Smith.
\newblock {Essays on Dynamic Models of Equilibrium and Learning.}
\newblock University of Chicago, 1991.

\bibitem[Xia and Conitzer(2010)]{XiaC10a}
Lirong Xia and Vincent Conitzer.
\newblock Stackelberg voting games: Computational aspects and paradoxes.
\newblock In \emph{Proceedings of the Twenty-Fourth {AAAI} Conference on
  Artificial Intelligence, {AAAI} 2010, Atlanta, Georgia, USA, July 11-15,
  2010}.

\bibitem[Xia et~al.(2011)Xia, Conitzer, and Lang]{XIACL11}
Lirong Xia, Vincent Conitzer, and J{\'{e}}r{\^{o}}me Lang.
\newblock Strategic sequential voting in multi-issue domains and
  multiple-election paradoxes.
\newblock In \emph{Proceedings of the 12th {ACM} Conference on Electronic
  Commerce (EC-2011), San Jose, CA, USA, June 5-9, 2011}, pages 179--188.

\end{thebibliography}
%\bibliographystyle{plain}
%\begin{verbatim}
%\documentclass[citeauthoryear]{llncs}
%\end{verbatim}
% \newpage
%\section{Appendix: missing proofs}
\end{document}